\documentclass[prl,twocolumn,floatfix,amsmath,amssymb]{revtex4}
\usepackage{epsfig}
\usepackage{graphicx}
\usepackage{txfonts}
\usepackage{natbib}

\begin{document}

\title{Hide and seek on complex networks}%

\author{Kim Sneppen${}^{1}$, Ala Trusina${}^{1,2}$, Martin Rosvall${}^{1,2}$}%
\affiliation{${}^{1}$Nordita, Blegdamsvej 17, 2100 Copenhagen {\O}}
\affiliation{${}^{2}$Department of Theoretical Physics, Ume{\aa} University}

\begin{abstract}
Signaling pathways and networks determine the ability to communicate
in systems ranging from living cells to human society.
We investigate how the network structure constrains communication in
social-, man-made and biological networks.
We find that human networks of governance and collaboration
are predictable on teat-a-teat level, reflecting well defined pathways, but globally inefficient.
In contrast, the Internet tends to have better overall communication abilities, more alternative pathways, and is therefore more robust.
Between these extremes the molecular network of
\emph{Saccharomyces cerevisea}e is more similar to the simpler
social systems, whereas the pattern of interactions in
the more complex \emph{Drosophilia melanogaster}, resembles the robust Internet.
\end{abstract}

\maketitle

Information exchange between distant parts of a complex
system is essential for its global functionality.
For example, without the adaptability to
environmental changes, maintained by communication through
signaling pathways, perturbations would be fatal for
living cells. Similarly, human society needs to
maintain global cooperativity in order to be functional. No
parts of such complex systems are complete, but all parts
are in contact with each other through a network of
distributed communication. The speed and reliability of the information
transfer is closely linked to the network architecture 
\citep{watts_strogatz,kleinberg,adamic,eckmann,dodds,rosvall}.
This interdependence can be characterized in terms of information measures
and Shannon entropies \citep{shannon}.
That is, we measure the number of
bits of information required to {\sl transmit} a message to a specific remote
part of the network (Fig.\ 1a), or reversely,
to {\sl predict} from where a message is received
(Fig.\ 1, b and c).
We will thus represent information measures related to
the network capacity for specific communication.
The introduced measures are not to be confused by 
the Shannon entropies that have
earlier been assigned to the network 
degree distribution \citep{sole},
respectively to the long time amplification of the dominant eigenvector
of the network adjacency matrix \citep{demetrius}.

\begin{figure}
\includegraphics[width=\columnwidth]{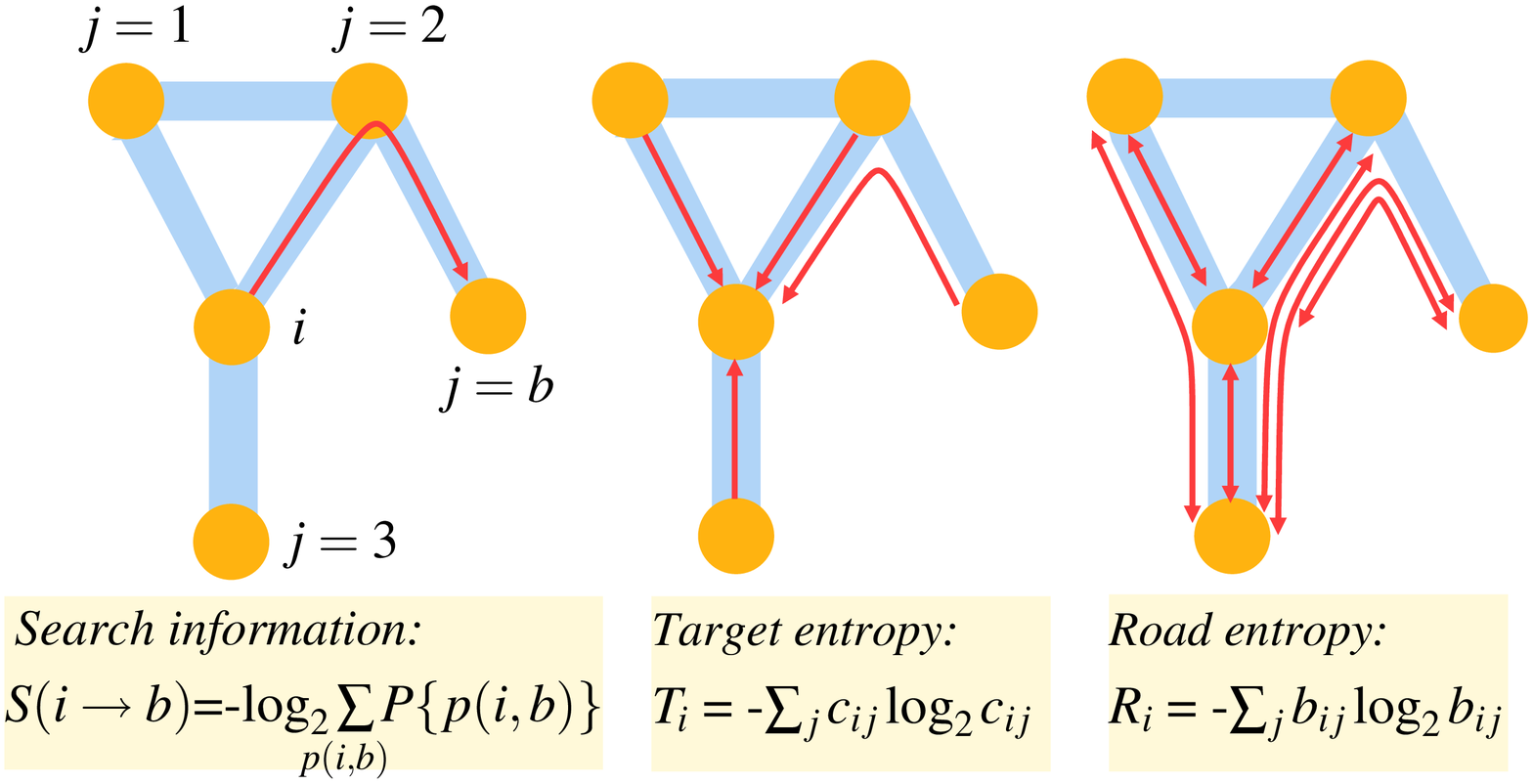}
\caption{Information measures on network topology:
{\sl \textbf{(a)}} {\it Search information} \mbox{$S(i \to b)$} measures your
ability to locate node $b$ from node $i$.
{\sl \textbf{(b)}} {\it Target entropy} $T_i$ measures
predictability of traffic
to you located at node $i$, and {\sl \textbf{(c)}}
{\it Road entropy} $R_i$ measures
predictability of traffic around $i$.
\mbox{$S(i \to b)$} is the number of yes/no questions needed to
locate any of the shortest paths between node $i$ and node $b$.
For each such path
$P\{p(i,b)\} \; =\; \frac{1}{k_i} \;\; \prod_{j}
\frac{1}{k_j-1}$, with $j$ counting nodes on the path
$p(i,b)$ until the last node before $b$. 
$c_{ij}$ is the fraction of the messages targeted to $i$ that passed
through neighbor node $j$. $b_{ij}$ is the fraction of
messages that go through node $i$ which also go through
neighbor node $j$. 
}
\label{sneppen_fig1}
\end{figure}

In practice, imagine that you at node $i$ want to send a message to node
$b$ in a given network (Fig.\ 1a).
This could for example correspond to sending an E-mail over the Internet.
For simplicity we assume that the message follow the shortest path,
or if there are several degenerate shortest paths, it is sent along one of them.
For each shortest path we calculate the probability
to follow this path, Fig.\ 1a, if one without 
information would chose any new direction with equal probability:
\begin{equation}
P\{p(i,b)\} \; =\; \frac{1}{k_i} \;\; \prod_{j\; \in\; p(i,b)}
\frac{1}{k_j-1},
\end{equation}
with $j$ counting all nodes on the path from a node $i$ to
until the last node before
the target node $b$ is reached. The factor $k_j-1$ instead of $k_j$
takes into account the information we gain
by following the path, and therefore reduce the number of exit links by one.
The total information needed to identify one of
all the degenerate paths between $i$ and $b$ 
defines the ``search information"
\begin{equation}
S(i\rightarrow b) = - log_2\left( \sum_{p(i,b)}  P\{ p(i,b) \} \right),
\end{equation}
where the sum runs over all degenerate paths that
connect $i$ with $b$. 
A large $S(i\rightarrow b)$
means that one needs many yes/no questions to locate $b$.
The existence of many degenerate paths will be 
reflected in a small $S$ and consequently in easy goal finding.

\begin{figure*}
\includegraphics[width=\textwidth]{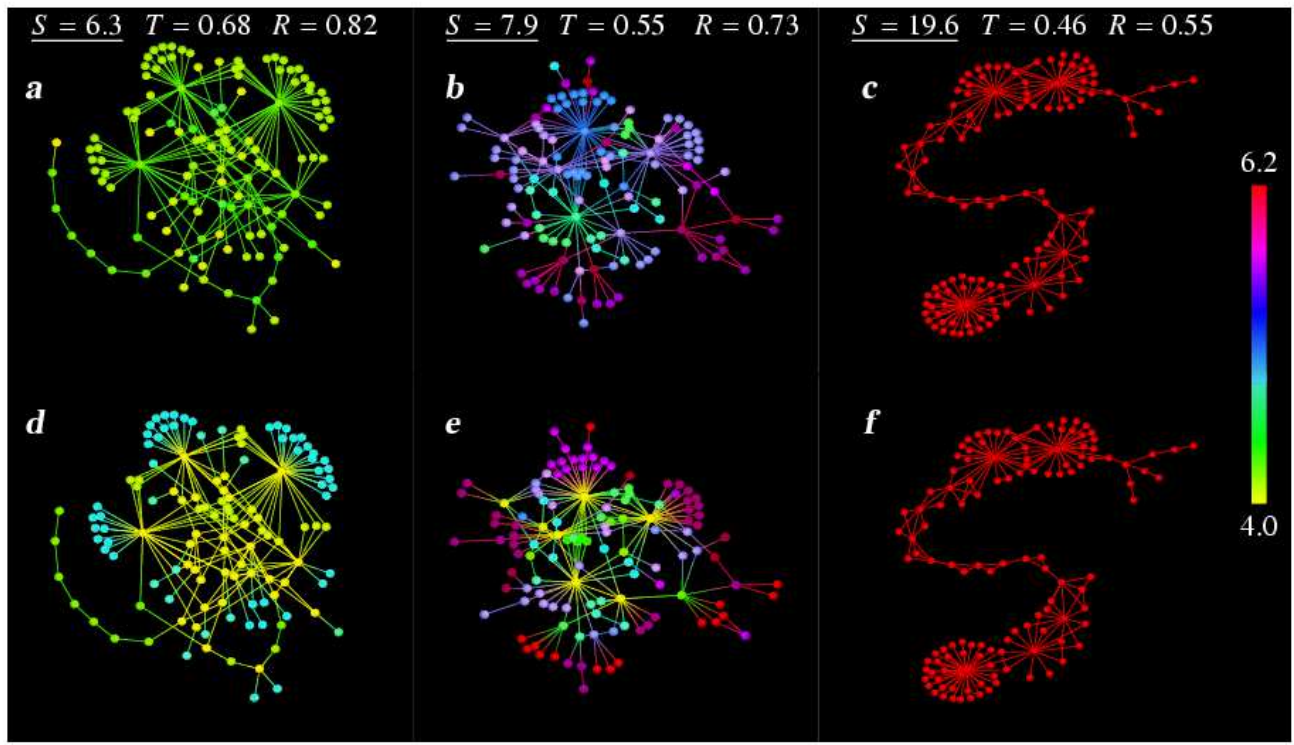}
\caption{Hide and seek in complex networks.
In {\sl \textbf{(a-c)}} we show two networks obtained by
{\sl \textbf{(a)}} minimizing and {\sl \textbf{(c)}} maximizing $S$,
 while keeping
the degree distribution identical to the Canadian hardwired
Internet in {\sl \textbf{(b)}}.
This network was selected as a typical communication network \citep{faloutsos,broder,barabasi},
with a broad degree distribution
$P(>k)\sim k^{-1.3}$.
The color of each node $i$ shows the value ${\cal A}_i
=\sum_b S(i\rightarrow b)$,
that measures how easy it is to find other nodes when starting at node $i$.
In {\sl \textbf{(d-f)}} we show same networks, but color coded according to
how difficult it is to find the nodes, $H_b=\sum_i S(i\rightarrow b)$.
}
\label{sneppen_fig2}
\end{figure*}

The practical question is thus:
Which position provides best access to the entire network?
Surfing the Web, which web-page should be the start page when
easy access to any other page is essential?
The answer is the node with minimal access information,
\mbox{${\cal A}_i=\sum_b S(i\rightarrow b)$}.
The networks in Fig.\ 2, a to c, are color coded
according to ${\cal A}_i$.
Fig.\ 2b illustrates
that hubs, and often nodes directly connected to hubs,
give best access to the system.
Overall one can see that it is easy to access other
nodes in the network in Fig.\ 2a,
whereas it is much more difficult in Fig.\ 2c.
In fact the network in Fig.\ 2b is the Canadian Internet \citep{internet},
whereas the networks in Fig.\ 2, a and c,
are obtained by rewiring the Canadian network to,
respectively, minimize and maximize
\mbox{$S=\sum_i {\cal A}_i/N$}
while maintaining the network connected and conserving
the degree of all nodes \citep{maslov2002}.
$N$ is the number of nodes in the connected network.

Naturally, the next question is:
Where it is best to hide?
That is where \mbox{${\cal H}_b= \sum_i S(i\rightarrow b)$} is maximal.
Note that maximizing everyone's ability to hide
$\sum_b {\cal H}_b = \sum_i {\cal A}_i = S \cdot N$
is equivalent to maximizing the search information
and therefore minimizing everybody's ability to search.
Thus we illustrate the value of ${\cal H}_b$
in Fig.\ 2, d to f, for the same networks as in Fig.\ 2, a to c.
In agreement with intuition we indeed find that hubs are easily accessible
by other nodes and thus are bad places for hiding.
Rather one should hide on nodes on the periphery.
Is it possible for a node to have a good access to other nodes
but not be easy accessible at the same time?
The compromise favors a position on a neighbor to a hub.
For example, if we consider the network implementation of
a city with roads as nodes and intersections as links,
it is preferable with an address on a small road that
connects directly to a major road/hub.

We will later see that many real world networks are characterized
by relatively high value of the overall search information $S$ (Fig.\ 4), implying that
global search abilities are limited by functional, geographical
or other constraints. The ability to search/hide is however not the only
measure of the communication properties of a network.
Another key aspect of communication handling is associated to
{\sl prediction} of
local traffic {\it to} and {\it across} nodes in the network.
This represents the ``passive" aspect of information handling.

To define the predictability, let us consider messages arriving
to a given node $i$ in a network. Your task, being on node $i$,
is to guess the ``active" neighbor/link from where the
next message arrives.
Without prior knowledge, all your local connections are equal
and it would take you $\log_2(k)$ yes/no questions to guess
the active link, where $k$ is the number of
connections of your node.
However, if the information about the
traffic through links is available,
the direction of the next message can be guessed
with less questions if the search
is biased towards the more used links.
For simplicity we assume
that all communication takes place through the shortest paths
and all nodes communicate in equal amounts with all other nodes.

\begin{figure*}
\includegraphics[width=\textwidth]{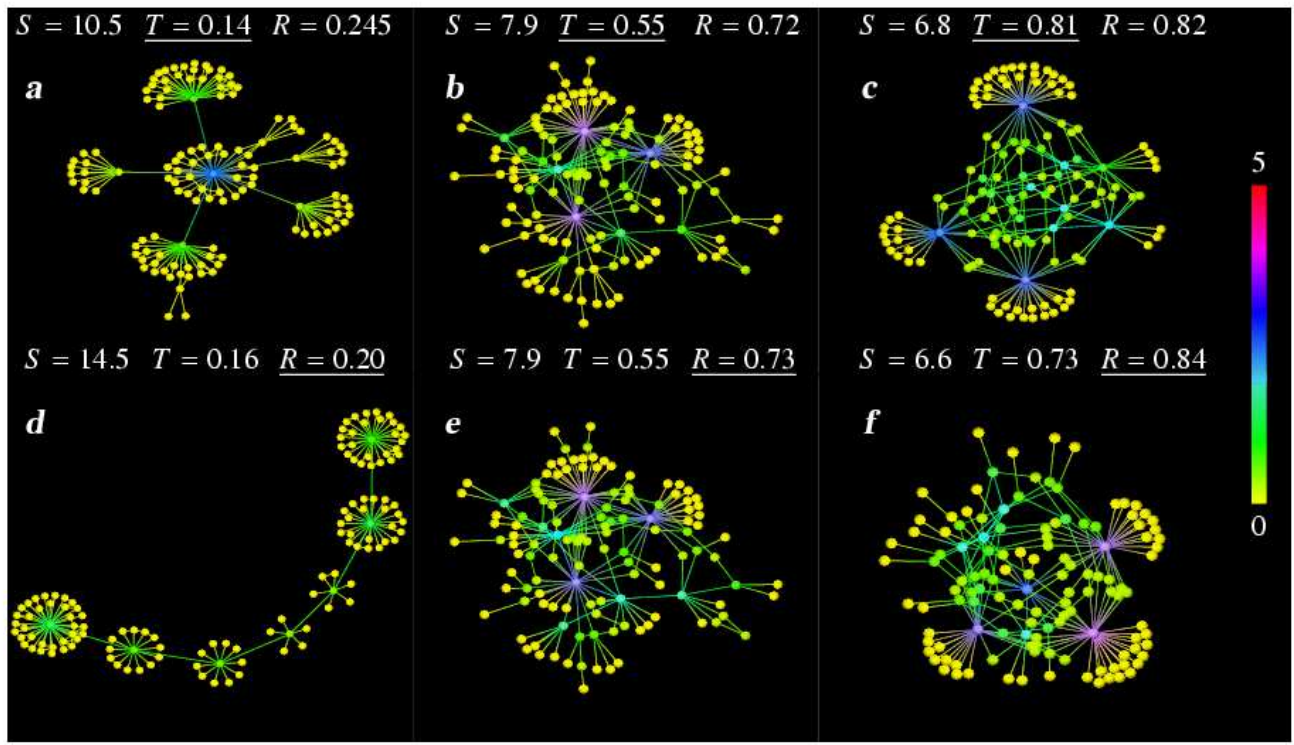}
\caption[]
{Prediction of local communication. The upper panel shows networks obtained by
(\textsl{\textbf{a}}) minimizing and (\textsl{\textbf{c}}) maximizing the target entropy $T= \sum_i T_i/N$ associated
to traffic {\it to} nodes in the networks, while keeping
the degree distribution identical \citep{maslov2002,maslov2002b}
to the original network of Autonomous Systems in Canada shown in (\textsl{\textbf{b}}).
(\textsl{\textbf{d}} to \textsl{\textbf{f}}) show the networks
that (\textsl{\textbf{d}}) minimize and (\textsl{\textbf{f}}) maximize the road entropy $R= \sum_i R_i/N$ associated
to traffic {\it across} nodes in the networks.
In (\textsl{\textbf{a}} to \textsl{\textbf{c}}) the nodes are color coded
according to the value of $T_i$, while
we in (\textsl{\textbf{d}} to \textsl{\textbf{f}}) color code according to $R_i$.
}
\label{sneppen_fig3}
\end{figure*}

The predictability, or alternatively the order/disorder of the
traffic around a given node $i$,
is measured by an entropy of messages that are targeted
{\sl to} a given node $i$, $T_i$, and an entropy
of all messages {\sl across} the node, $R_i$ (Fig. 1, b and c).
The predictability based on the orders that are targeted
{\sl to} a given node $i$ is
\begin{equation}
T_i \; = \; - \sum_{j=1}^{k_i} c_{ij} log_2(c_{ij}),
\end{equation}
where $j=1,2...,k_i$ denotes the links from node $i$
to its immediate neighbors $j$ and
$c_{ij}$ is the fraction of the messages targeted to $i$
that passed through node $j$.
Similarly we use
$b_{ij}$, defined as the fraction of messages that
go through node $i$ that also go through node $j$,
to quantify the entropy associated to traffic {\sl across}
node $j$:
\begin{equation}
R_i \; = \; - \sum_{j=1}^{k_i} b_{ij} log_2(b_{ij}),
\end{equation}
Technically $b_{ij}$ is proportional to the betweenness
\citep{between} 
of the link between $i$ and $j$, whereas $c_{ij}$ rather
quantifies a sub-division of the network around node $i$.
We will refer to $T_i$ as the target entropy, and to 
$R_i$ as the road entropy, where
a large $T_i$ or $R_i$ mean a low predictability.

Fig.\ 3 shows the values of
$T_i$ and $R_i$ for different complex networks.
In Fig.\ 3, a to c, we examine networks by color coding
the nodes according to target entropy, $T_i$.
Fig.\ 3, d to f, show networks color coded
according to the road entropy $R_i$.
The bluish hubs reflect that
traffic to highly connected nodes is hard to predict.
However, this is not always the case: the location
of nodes with low predictability also depends
on the overall topology of the network.
The networks in Fig.\ 3 are presented so that
the entropy increases from, respectively, a and d to c and f.
As the networks get more disorganized, the number of hubs
with disordered traffic increases. Also,
nodes of low degree become more confused as they tend to
position themselves between the hubs.
It is interesting that this positioning of low degree nodes
increases the number of alternative pathways in the system,
and thus tend to minimize the search information $S$.
Therefore the minimal $S$ network in Fig.\ 2a
is similar to the maximal $R=\sum_i R_i$ or
$T=\sum_i T_i$ networks in Fig.\ 3, c and f.

Whereas the maximal $T$ and $R$ networks are topologically similar,
this is not at all the case for the minimal $T$ and $R$ networks in Fig.\ 3, a and d.
The network of minimal $T$ in Fig.\ 3a concentrates all signaling into
a simple star like structure with hierarchical features \citep{trusina2004}.
As a consequence nearly everybody can easily predict
from where the next message will come.
In contrast, minimizing $R$ results in a
topology characterized by hubs on a string forming an
``information super highway'' (Fig.\ 3d).
Thus a low road entropy $R$ means that relatively many links are important,
whereas a large $R$ implies that few links are essential.
In this sense $R$ is related to
robustness in an intentional edge attack \citep{newman2003}
whereas $T$ reflects robustness in an intentional node attack \citep{newman2003}.

We apply our information measures to characterize real networks
in Fig.\ 4, by comparing a number of networks with their
randomized counterparts \citep{maslov2002}.
The datails of the comparison is shown in table 1.
For each network we show the Z-score for
$S$, $T$ and $R$. A large positive $Z$-score
means that the corresponding
network has relatively large entropy.
For example we see that the hardwired Internet is quite ``messy" in all
senses: The traffic is unpredictable,
implying that the network is robust, and at the same time
one needs relatively large information handling to transmit
packages across the system.
In contrast the social networks,
exemplified here by the network of company executives in USA, CEO \citep{ceo}
and the scientific collaboration network, hep-th \citep{th-hep},
show a pronounced pattern of high traffic predictability and large
cost of locating any particular node. These features are
characteristic to the ordered network topologies in
Fig.\ 3a and d.

\begin{figure}
\includegraphics[width=\columnwidth]{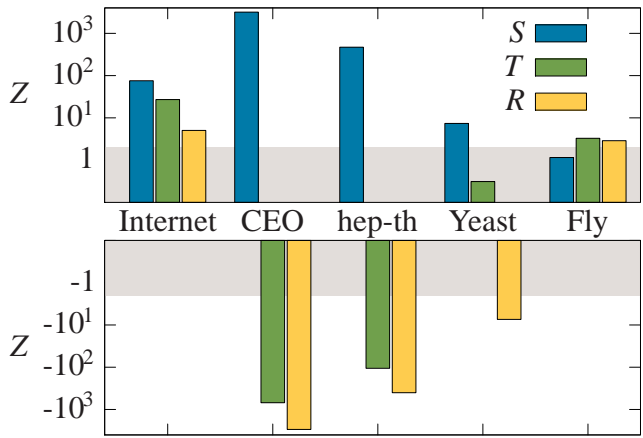}
\caption{Measure of relative order in
communication networks.
A high $Z$-score implies relatively high entropy.
In all cases we show $Z = (I-I_r)/\sigma_r$ for
the information measures $I=$ $S$, $T$ and $R$,
by comparing with $I_r$ for randomized networks
with preserved degree distribution. $\sigma_r$ is the standard deviation
of the corresponding $I_r$, sampled over 100 realizations.
Results within the shaded area of two standard deviations are insignificant.
All networks have a relatively high search information $S$.
The two human interaction networks CEO \citep{ceo} and
scientific collaborations \citep{th-hep} show
a distinct communication structure characterized by local
predictability, low $T$ and $R$, and global inefficiency, high $S$.
}
\label{sneppen_fig4}
\end{figure}

In Fig.\ 4 we also investigate networks of physical interactions
among proteins in two organisms, yeast \citep{Uetz2000,Ito2001}
and fly \citep{giot}.
Whereas the fly network is quite close to its randomized counterpart,
yeast is reminiscent of the social networks.
The large $S$ for yeast
reflects that many of the largest hubs are positioned on the
periphery of the network \citep{maslov2002},
and therefore have relatively large entropy ${\cal A}_i$, see Fig.\ 5.
This tendency of hub separation reflects optimization of local communication,
at the cost of global specific signaling.
On the other hand the protein network of the multicellular and more advanced
fly, \emph{Drosophilia melanogaster}, displays a more complicated and in fact
more robust topology as witnessed by the significantly positive
$Z-$scores for $T$ and $R$ entropies.

\begin{table}
\caption{\label{table}
Measure of order in
communication networks. 
Five networks together with their size $N$ and the three
information-entropy measures $I=S$, $T$ and $R$. 
In each case we compare the measured $I$-value
by comparing with $I_r$ for randomized networks
with preserved degree distribution. $\sigma_r$ is the standard deviation
of the corresponding $I_r$ sampled over 100 realizations.
For all networks we only consider the largest connected component,
that is also maintained during the randomization.
The Internet network is the hardwired Internet 
of autonomous systems \citep{internet}.
The CEO network is chief executive officers connected by links when
they sit in the same board of directors \citep{ceo},
hep-th is a network of scientists connected by links if they
coauthor a publication \citep{th-hep},
yeast is the protein-protein interaction network
in \emph{Saccharomyces Cerevisiae} 
detected by the two-hybrid experiment\citep{Uetz2000},
and fly refers to the similar network
in \emph{Drosophilia melanogaster} \citep{giot}.
Both of these networks are pruned to include only interactions
of high confidence, and in both networks we compare with their
random counterparts where both bait and prey connectivity of all proteins
are preserved.
The results on the network of \citep{Uetz2000}
is reproduced when considering the core of the yeast network measured
by \citep{Ito2001}.
Furthermore, all results are robust to a 10\% random removal of links
except for the fly network which with such a pruning tends to
be closer to the yeast network.
}
\label{table:realNets}
\centering
\begin{tabular}{l|r|r@{.}l|r@{.}l|r@{.}l|}
\multicolumn{1}{c}{Network} &
\multicolumn{1}{c}{$N$} &
\multicolumn{2}{c}{$S$} &
\multicolumn{2}{c}{$T$} &
\multicolumn{2}{c}{$R$}\\
\hline
\hline
Internet & $6474$ & $16$&$34$ & $0$&$583$ & $0$&$809$\\
{\small randomized} & $$ & $15$&$03(2)$ & $0$&$499(3)$ & $0$&$793(3)$\\
\hline
CEO & $6193$ & $20$&$693$ & $1$&$58$ & $1$&$831$\\
{\small randomized} & $$ & $12$&$597(3)$ & $3$&$316(3)$ & $3$&$513(1)$\\
\hline
hep-th & $5835$ & $19$&$72$ & $0$&$847$ & $1$&$211$\\
{\small randomized} & $$ & $13$&$48(1)$ & $1$&$385(5)$ & $1$&$668(1)$\\
\hline
Yeast \citep{Uetz2000} & $921$ & $13$&$3$ & $0$&$38$ & $0$&$722$\\
{\small randomized} & $$ & $12$&$5(1)$ & $0$&$38(1)$ & $0$&$742(3)$\\
Yeast \citep{Ito2001} & $417$ & $12$&$2$ & $0$&$30$ & $0$&$662$\\
{\small randomized} & $$ & $10$&$7(2)$ & $0$&$33(2)$ & $0$&$708(6)$\\
\hline
Fly & $2915$ & $14$&$03$ & $0$&$56$ & $0$&$931$\\
{\small randomized} & $$ & $13$&$96(6)$ & $0$&$53(1)$ & $0$&$925(2)$\\
\hline
\end{tabular}
\end{table}

\begin{figure}
\includegraphics[width=\columnwidth]{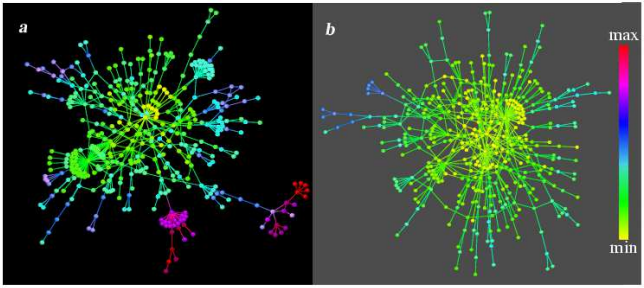}
\caption{
Analysis of the protein-protein interaction network in yeast:
{\sl \textbf{(a)}} shows the core of the yeast protein-protein interaction
network color coded according to ${\cal A}_i$.
The core data is the reliable subset of the two hybrid data set of
Ito et\ al. \citep{Ito2001}, obtained by selecting the largest
connected component of the network with only interactions with at least 3 IST
included.
The value of ${\cal A}_i$ is increasing from yellow through green and blue
to red
which mark nodes that have least access to the rest of the network.
Proteins colored in red are proteins with ``specific signaling pattern''.
{\sl \textbf{(b)}} shows an example of a randomized version of the network in
{\sl \textbf{(a)}},
also color coded with the corresponding ${\cal A}_i$ values.
The network is randomized such that the number of links for each
node (protein) is maintained, while interaction partners are
reshuffled in the same way as in the analysis leading to Fig.\ 4.
The access information are substantially lower than in {\sl \textbf{(a)}},
reflecting a topology with better global access at the cost
of higher possibility to be disturbed.
}
\label{sneppen_fig5}
\end{figure}

Networks are inherently coupled to communication and indeed
their topology reflects this. The optimal
topology for information transfer relies on a system-specific
balance between effective communication (\emph{search}) and
not having the individual parts being unnecessarily disturbed (\emph{hide}).
We have presented measures that quantify the ease of global
search, $S$, and the predictability of local activity, $T$ and $R$,
and illustrated how they characterize the organization of complex networks.

In particular the network of corporate CEOs and
scientific co-authorship,
were found to be highly ``predictable", and at the same time very
inefficient in transmitting information.
In contrast the hardwired Internet was found to
be locally unpredictable, and therefore robust against local failures.
Finally the fruit fly, \emph{Drosophilia melanogaster},
 has a more robust protein network
than yeast, \emph{Saccromyces cerevisiae},
with better connections between distant parts of the network.
This global communication optimization may reflect that the
multicellular organism must sustain life in cells with many
more different local environments than the single-celled yeast.
\\\\

We thank R. Donangelo, P. Minnhagen and J. Wakeling
for useful comments and revisions of the manuscript.
We acknowledges the support of
Swedish Research Council through Grants No. 621 2003 6290
and 629 2002 6258 and of EU through the Evergrow project.\\
Correspondence and requests for materials
should be addressed to K. Sneppen~(email: sneppen@nbi.dk).

\end{document}